\documentstyle[here,epsfig]{mn2e}
\newcommand{\beq}{\begin{equation}}
\newcommand{\eeq}{\end{equation}}
\newcommand{\bdm}{\begin{displaymath}}
\newcommand{\edm}{\end{displaymath}}
\newcommand{\bfig}{\begin{figure}}
\newcommand{\efig}{\end{figure}}
\newcommand{\msun}{M_{\odot}}
\def\etal{{ et al.~}}
\def\then{$\rightarrow$}
\def\ie{{\frenchspacing\it i.e. }}
\def\eg{{\frenchspacing\it e.g. }}

\def\gtsima{$\; \buildrel > \over \sim \;$}
\def\ltsima{$\; \buildrel < \over \sim \;$}
\def\prosima{$\; \buildrel \propto \over \sim \;$}
\def\gsim{\lower.5ex\hbox{\gtsima}}
\def\lsim{\lower.5ex\hbox{\ltsima}}
\def\simgt{\lower.5ex\hbox{\gtsima}}
\def\simlt{\lower.5ex\hbox{\ltsima}}
\def\simpr{\lower.5ex\hbox{\prosima}}

\def\HH{H$_2$~}

\title[HD cooling in primordial star formation]{The role of HD cooling
  in primordial star formation}

\author[Emanuele Ripamonti]{Emanuele Ripamonti$^{1}$\thanks{E-mail:
ripa@astro.rug.nl}\footnotemark[1]\\
$^{1}$ University of Groningen, Kapteyn Astronomical Institute, Postbus 800,
9700 AV Groningen, The Netherlands}
\begin{document}

\date{Submitted: June 2006; accepted: January 2007}

\pagerange{\pageref{firstpage}--\pageref{lastpage}} \pubyear{2007}

\maketitle

\label{firstpage}

\begin{abstract}
The role of HD cooling in the formation of primordial objects is
examined by means of a great number of 1-D models of the collapse of
halos, exploring a wide range of masses and virialization redshifts. We
find that HD has very little effect upon the critical mass separating
the objects which are likely to form stars from those which are not. We
also find that, once the proto-stellar collapse has started, HD effects
are quite negligible.

Instead, HD effects can be important during the intermediate stage of
gas fragmentation: objects below a certain mass scale ($\sim 3\times
10^5\msun$ at $z_{\rm vir}=20$ in our ``fiducial'' case) can be cooled
by HD down to $T\sim50-100\;{\rm K}$, whereas \HH cooling never takes
the gas below $T\sim 200\;{\rm K}$.  The lower temperature implies a
reduction of a factor $\sim 10$ in the Jeans mass of the fragmenting
gas, and stars forming in such low-mass halos are probably less massive
than their counterparts in larger halos. We estimate the importance of
this mode of star formation through a variation of the
Press-Schechter formalism, and find that it never exceeds the
contribution of halos which are cooled by \HH only. Halos where HD is
important account at best for a fraction $\sim$0.25 of the total
primordial star formation.  However, HD cooling might provide a channel
through which long-lived low mass stars could be formed in primordial
conditions.
\end{abstract}

\begin{keywords}
stars: formation -- molecular processes -- cosmology: theory.
\end{keywords}

\section{Introduction}


Molecules play a decisive role in the formation of the first generations
of luminous objects, because they provide the main cooling mechanism for
primordial metal-free gas in small halos virializing at high
redshift (\eg\ Barkana \& Loeb 2001; Bromm \& Larson 2004; Glover 2004;
Ciardi \& Ferrara 2005; Ripamonti \& Abel 2005).

The \HH molecule has long been recognized (Saslaw \& Zipoy 1967, Peebles
\& Dicke 1968, Matsuda, Sato \& Takeda 1969) as the most important
cooling agent in such conditions. Its cooling and chemical
properties have been carefully studied (\eg Hollenbach \& McKee 1979;
Palla, Salpeter \& Stahler 1983 [PSS83]; Stancil, Lepp \& Dalgarno
1996,1998; Abel \etal 1997 [A97]; Galli \& Palla 1998 [GP98], Glover,
Savin \& Jappsen 2006; Hirata \& Padmanabhan 2006) and are a
key component of all recent theoretical models and numerical
investigations of the formation of the first objects (\eg\ Tegmark \etal
1997 [T97]; Omukai \& Nishi 1998; Bromm, Coppi \& Larson 1999 [BCL99], 2002
[BCL02]; Abel, Bryan \& Norman 2000, 2002 [ABN02]; Ripamonti
\etal 2002 [R02], Yoshida \etal 2006 [Y06]).

Such studies suggest that primordial stars are more
massive than their present-day counterparts, and that this difference
arises from the different cooling properties of metal-free and
metal-enriched gas (Bromm \etal 2001; Omukai \& Palla 2001, 2003;
Schneider \etal 2002). A direct link seems to exist between the
properties of \HH cooling and the large fragment mass and the high
proto-stellar accretion rate which are found to occurr in simulations of
primordial objects.

However, most of these studies consider \HH as the only molecular
coolant of primordial gas. This assumption might be quite critical,
because of the close link between molecular and stellar properties. In
fact, a number of chemical models of the primordial medium (Puy \etal
1993; GP98; Stancil, Lepp \& Dalgarno 1998) have actually shown that, if
the primordial gas cools below $\sim 200\;{\rm K}$, then HD molecules
are likely to become the main cooling agents, despite their low number
abundance. The cooling properties of HD and \HH are quite different from
each other, so this could have important consequences: for instance, in
an HD-cooled gas, the Jeans mass (and the typical mass of primordial
objects) would likely be reduced.

Such a possibly important role prompted further investigations of the HD
cooling and chemistry (Flower 2000; Galli \& Palla 2002 [GP02]; Lipovka,
Nu\~nez-Lopez \& Avila-Reese 2005) and especially of its effects on the
properties of the first generations of stars (BCL99, BCL02; Uehara \&
Inutsuka 2000; Flower \& Pineau des For\^ets 2001; Nakamura \& Umemura
2002; Johnson \& Bromm 2006; Lipovka \etal 2005; Nagakura \& Omukai
2005; Shchekinov \& Vasiliev 2006; Y06).

All the studies which find that HD effects might be important were
considering situations where the primordial gas was already
``perturbed'' by some kind of phenomenon, such as the shock due to the
merger of two halos, or the radiation from the first stars.  The
``unperturbed'' case was considered by the detailed simulations by
BCL99, BCL02, and Y06, who included a treatment of HD chemistry and
cooling in some of their simulations, and found that HD had minor
effects at best. However, such results concern the evolution of single
``typical'' cases and do not rule out the existence of objects where HD
cooling is actually important.

In this paper, we address the question of whether HD can be
important during the birth of the {\it first generation} of luminous
objects (that is, in {\it unperturbed} gas), with regard to the
properties of the halos where they form, to the properties of
fragmentation inside such halos, and to the characteristics of the
proto-stellar collapse. This is done by means of spherically symmetric 1D
calculations of the evolution of gas properties during the collapse of
halos or proto-stars, through which we carry out an exploration of a
large portion of the $z_{\rm vir}-M_{\rm halo}$ parameter space. Such
exploration, combined with results from analytical models, allows us to
give a quantitative estimate of the importance of HD-cooled objects.

This paper is organized as follows: in section 2 we describe the details
of the 1D calculations; in section 3 we examine the results obtained
from such calculations, and in section 4 we discuss their cosmological
significance. Finally, in section 5 we summarize our conclusions.

Throughout all of this paper, we assume a flat
$\Lambda$CDM cosmological model with parameters taken from the three
years WMAP results reported by Spergel \etal 2006:
$\Omega_\Lambda=0.76,\ \Omega_{\rm m}=0.24,\ \Omega_{\rm b}=0.042,\
\Omega_{\rm DM}\equiv\Omega_{\rm m}-\Omega_{\rm b}=0.198,\ h=0.73$.
Furthermore, $\rho_0\simeq 1.88\times
10^{-29} h^2\;{\rm g\,cm^{-3}}$ is the critical density of the universe,

\section{Method}

In order to assess the effects of HD in a wide range of cosmological
environments, a method similar to that of T97 is adopted: rather than
simulating a single ``typical'' object in great detail (as was done by
BCL99, BCL02 and Y06), the ``coarse'' behaviour of a plethora of halos
is probed, exploring the $z_{\rm vir}-M_{\rm halo}$ parameter space.


\subsection{The 1D Code}

\begin{table}
\caption{List of considered reactions}
\label{reactions_list}
\begin{tabular}{crcll}
\hline\hline
\# & Reaction & & & Reference\\
\hline
 1 & H$^+$ + $e^-$    &\then& H + $\gamma$         & A97/2$^{a}$\\
   &                  &     &                      & GP98/H1$^{a,b}$\\
 2 & H + $\gamma$     &\then& H$^+$ + $e^-$        & GP98/H2$^{b}$\\
 3 & H + $e^-$        &\then& H$^-$ + $\gamma$     & A97/7\\
 4 & H$^-$ + $\gamma$ &\then& H + $e^-$            & GP98/H4\\
 5 & H + H$^-$        &\then& \HH + $e^-$          & A97/8\\
 6 & H$^-$ + H$^+$    &\then& H + H                & A97/16\\
 7 & H + H$^+$        &\then& H$_2^+$+ $\gamma$    & GP98/H8\\
 8 & H$_2^+$+ $\gamma$&\then& H + H$^+$            & GP98/H9\\
 9 & H$_2^+$+ H       &\then& \HH + H$^+$          & A97/10\\
10 & \HH + H$^+$      &\then& H$_2^+$+ H           & GP98/H15\\
11 & D$^+$ + $e^-$    &\then& D + $\gamma$         & A97/2$^{a}$\\
   &                  &     &                      & GP98/D1$^{a,b}$\\
12 & D + $\gamma$     &\then& D$^+$ + $e^-$        & GP98/D2$^{b}$\\
13 & H$^+$ + D        &\then& H + D$^+$            & GP02/5\\
14 & H + D$^+$        &\then& H$^+$ + D            & GP02/6$^{c}$\\
15 & D$^+$ + \HH      &\then& HD + H$^+$           & GP02/2\\
16 & HD + H$^+$       &\then& D$^+$ + \HH          & GP02/4\\
17 & H + $e^-$        &\then& H$^+$ + $e^-$ + $e^-$& A97/1\\
18 & H + H            &\then& H$^+$ + H + $e^-$& PSS83/9\\
19 & He + $e^-$       &\then& He$^+$ + $e^-$ + $e^-$& A97/3\\
20 & He$^+$ + $e^-$   &\then& He + $\gamma$    & A97/4\\
21 & He$^+$ + $e^-$   &\then& He$^{++}$ + $e^-$ + $e^-$& A97/5\\
22 & He$^{++}$+ $e^-$ &\then& He$^+$ + $\gamma$& A97/6\\
23 & D + \HH          &\then& HD + H           & GP02/1$^{d}$\\
24 & HD + H           &\then& D + \HH          & GP02/3$^{d}$\\
25 & H + H + H        &\then& \HH + H          & PSS83/4\\
26 & \HH + H          &\then& H + H + H        & PSS83/5\\
27 & H + H + \HH      &\then& \HH + \HH        & PSS83/6\\
28 & \HH + \HH        &\then& H + H + \HH      & PSS83/7\\
\hline\hline
\end{tabular}
\\
The reference codes give the paper
from which the reaction coefficient was taken, and the number
identifying each reaction in the corresponding paper.\\
$^{a}$ We adopted the GP98 (case B) recombination rate before the
turn-around redshift of each halo,  when we switched to the A97 (case A)
rate.\\
$^{b}$ The coefficients given by GP98 for the ionization of H (and D)
were transformed into reaction rates we could promptly use through
eq. (1) of Sasaki \& Takahara 1993.\\
$^{c}$ As noted by Johnson \& Bromm 2006, the reaction coefficient given
by GP02 contains a typo; therefore, we actually use the original rate of
Savin (2002).\\
$^{d}$ When extrapolated to low temperature ($\lsim 100\;{\rm K}$),
these reaction rates become extremely large; since this is an artifact
of the adopted fitting forms, and there is no experimental results at
$T< 170\;{\rm K}$, we assume that at $T\leq100\;{\rm K}$ the rate
coefficients of these reactions remain constant at the value we obtain
for $T=100\;{\rm K}$.
\end{table}

The main tool employed in this paper is a 1D Lagrangian hydrodynamical
spherically symmetric code, which follows the evolution of primordial
gas inside and around DM halos, from the recombination epoch until well
after the DM has virialized. Such code is described in R02 (see also
Thoul \& Weinberg 1995; and Omukai \& Nishi 1998). Here we just outline
the changes that were made in order to adapt it to our present purposes.

\subsubsection{Chemistry}

The chemical network of the original R02 code was revised and
expanded. Now it follows the evolution of 12 species (the 9 ``original''
species, \ie H, H$^+$, H$^-$, \HH, H$_2^+$, He, He$^+$, He$^{++}$, and
$e^-$; plus D, D$^+$, and HD). The considered reactions are listed in
Table \ref{reactions_list}. The network includes all the reactions for
hydrogen and deuterium species which are in the {\it minimal model} of
GP98, the collisional ionizations of hydrogen and helium, and two other
deuterium reactions described by GP02 (these reactions were not part of
the GP98 minimal model). Finally, there are the 3-body reactions of \HH
formation from PSS83. We stress that reactions 3 and 5 (13 and 15) are
the main formation channel for \HH (HD), unless the density is so high
($\rho\gsim10^{-16}\,{\rm g\,cm^{-3}}$) that 3-body reactions (25 and
27) dominate.

As in the original code, the non-equilibrium chemistry of the considered
species is solved at each time step through an implicit difference
scheme.

It should be pointed out that, even if the rates given in Table
\ref{reactions_list} are fairly updated, their choice remains quite
slippery: the uncertainties range from 10-20 per cent up to one order of
magnitude in the worst cases (reactions 3, 4, 5, and 6, including the
main formation channel of \HH). Glover \etal (2006) and Hirata \&
Padmanabhan (2006) find that the large uncertainties in these ill known
reaction rates can have important effects on cosmological predictions;
but also that the formation of the first protogalaxies from cold,
neutral gas is relatively insensitive to the choice of the rates for
these reactions.  If so, the most problematic rate in our chemistry
calculations is likely to be the one for reaction 15 (which is part of
the main HD formation route): the value given by GP02 agrees within
20-30 per cent with that from Wang \& Stancil (2002), but there exist
measurements differing by almost a factor of 2 from these theoretical
estimates (see fig. 3 of GP02, and fig. 14 of Wang \& Stancil 2002). We
will try to estimate the effects of this uncertainty by running a
dedicated set of simulations (see Section 2.2.2).

It should also be noted that the rates of all the reactions which are
caused by the interaction with a photon (2, 4, 8, and 12) only keep into
account the effects of the CMB, while neglecting any extra radiation
field which might be present (see Hirata \& Padmanabhan 2006 for a
possible effect of extra radiation).

\subsubsection{Cooling}

The original code included the treatment of cooling from \HH
roto-vibrational lines (including their radiative transfer) and from the
CIE (Collision-Induced Emission) continuum. This was subject to several
changes:
\begin{enumerate}
\item{The treatment of \HH cooling was simplified in order to make the
  code faster: for this reason, we used the \HH cooling rate from GP98,
  and treated the effects of line radiative transfer by using the
  methods devised in Ripamonti \& Abel (2004) (in particular, of their
  equation 22); such methods are quite approximate, because the effects
  of line transfer are estimated only from the local density, whereas
  they depend also on temperature and velocity gradients. But they are
  adequate for the purposes of this paper.}
\item{The treatment of CIE cooling was switched off, as we never
  approach the density regime where it is important ($n\gsim
  10^{14}\;{\rm cm^{-3}}$).}
\item{We included the cooling due to HD molecules, adopting the results
  of Lipovka \etal (2005), which consider also the effect of vibrational
  lines, and are valid in a wide range of temperatures ($40\;{\rm K}\leq
  T\leq 2\times 10^4\;{\rm K}$) and densities ($1\; {\rm cm^{-3}}\leq
  n_{\rm H} \leq 10^8\;{\rm cm^{-3}}$; this range can be easily
  extrapolated to both higher and lower densities)\footnote{We actually
  used the polynomial fit provided by Lipovka \etal (2005); the fit is
  quoted to be close to their original results for $T\geq 100\;{\rm K}$;
  however, it is reasonably accurate also for $40\;{\rm K} \leq T \leq
  100\;{\rm K}$.}.}
\item{As we are concerned with a cosmological scenario, we included
  the cooling (or heating) from Compton scattering of CMB photons, and
  from Lyman $\alpha$ line cooling of atomic hydrogen. The adopted
  cooling rates per unit volume (from Rybicki \& Lightman 1979 and
  Dalgarno \& McCray 1972, respectively) are
  \begin{eqnarray}
    \Lambda_{\rm C}(T,T_\gamma) & \simeq & 1.0 \times 10^{-37}
    n_{e} T_\gamma^4 (T_\gamma-T) \; {\rm erg\,s^{-1}\,cm^{-3}}\\
    \Lambda_{\rm H}(T) & \simeq & 7.5 \times 10^{-19} e^{-118348/T} n_e
    n_{\rm H} \; {\rm erg\,s^{-1}\,cm^{-3}}
  \end{eqnarray}
  where $T$, $T_\gamma$ ($\simeq 2.725(1+z)$), $n_e$ and $n_{\rm H}$ are
  the temperatures of the gas and of the cosmic background radiation
  (both in K), and the number densities of free electrons and H atoms
  (both in ${\rm cm^{-3}}$), respectively.}
\item{When the temperature of the gas is of the same order as that of
  the cosmic background radiation, the cooling rates from \HH and HD can
  be substantially modified (see \eg fig. 9 of GP02), as atoms and
  molecules become {\it heating} agents when $T<T_\gamma$. We
  account for the effects of the CMB upon the \HH and HD cooling rates
  by estimating the {\it net} cooling rates through an adaptation of
  eq. (6) of Puy \etal (1993)\footnote{In the Puy \etal (1993) paper,
  the sign of the argument of the exponential is wrong because of a
  typografical error.}
\begin{eqnarray}
\Lambda_{{\rm net},H_2}(T,T_\gamma) \simeq \Lambda_{\rm H_2}(T) [1 -
  e^{T_{\rm H_2}\left({{1\over T}-{1\over T_\gamma}}\right)}].\\
\Lambda_{{\rm net,HD},j}(T,T_\gamma) \simeq \Lambda_{\rm HD}(T) [1 -
  e^{T_{\rm HD}\left({{1\over T}-{1\over T_\gamma}}\right)}].
\end{eqnarray}
  where $T_{\rm H_2}\simeq 512\;{\rm K}\;(T_{\rm HD}\simeq128\;{\rm K})$
  is the excitation temperature of $H_2$ (HD). The CMB also affects
  the Lyman $\alpha$ cooling, but this is completely negligible, as
  $\Lambda_{\rm H}(T)$ is extremely small for $T\sim T_\gamma$.}
\end{enumerate}

\subsubsection{Dark Matter}
\label{subsubsection_darkmatter}


The gravitational effects of Dark Matter (DM) were introduced in the
code. The DM halo was assumed to be always spherically symmetric and
concentric with the simulated region, whose central part represents the
investigated halo. A DM mass of $M_{\rm DM} = M_{\rm halo}
\Omega_{\rm DM} / \Omega_{\rm m}$ is assumed to be within a certain
truncation radius $R_{\rm tr}$, inside which the DM density profile is a
truncated isothermal sphere with a flat core of radius $R_{\rm core}$;
outside the truncation radius, the DM density is assumed equal to the
cosmological average. So,
\begin{equation}
\rho_{\rm DM}(r) =
      	\left\{{
	\begin{array}{ll} 
	\rho_{\rm core}
	    & {\rm if}\ r \leq R_{\rm core};\\
	\rho_{\rm core} (r/R_{\rm core})^{-2}
	    & {\rm if}\ R_{\rm core} \leq r \leq R_{\rm tr};\\
	\rho_0 \Omega_{\rm DM} (1+z)^3
	    & {\rm if}\ r \ge R_{\rm tr},\\
        \end{array}}\right .
\label{DM_density_profile}
\end{equation}
where the core density $\rho_{\rm core}$ is
\begin{equation}
\rho_{\rm core} = M_{\rm DM} \left[{ {{4\pi}\over3} R_{\rm core}^3
    \left({{{3R_{\rm tr}}\over{R_{\rm core}}} -
    2}\right)}\right]^{-1},
\label{rhocore_evolution}
\end{equation}
as can be obtained by equating the DM mass within $R_{\rm tr}$ to
$M_{\rm DM}$.

The truncation and core radii were assumed to evolve in time, mimicking
the results for the evolution of a simple {\it top-hat} fluctuation (see
\eg\ Padmanabhan 1993), and are described by the following equations
\begin{equation}
  R_{\rm tr}(z) = 
  \left\{{
    \begin{array}{ll}
      \left({{3\over{4\pi}}
	{{M_{\rm DM}}\over{\rho_{\rm TH}(z)}}}\right)^{1/3}
      & {\rm if\ }z \geq z_{\rm ta}\\
      R_{\rm vir}
      \left[{2-{{t(z)}\over{t(z_{\rm vir})-t(z_{\rm ta})}}}\right]
      & {\rm if}\ z_{\rm ta} > z \ge z_{\rm vir}\\
      R_{\rm vir}
      & {\rm if}\ z < z_{\rm vir}\\
  \end{array}}\right .
\end{equation}
\begin{equation}
  R_{\rm core}(z) = 
  \left\{{
    \begin{array}{ll} 
      R_{\rm tr}(z) & {\rm if\ } z \ge z_{\rm ta}\\
      R_{\rm vir}
      \left[{2-
	       {{(2-\xi) t(z)}\over{t(z_{\rm vir})-t(z_{\rm ta})}}}\right]
      & {\rm if\ } z_{\rm ta} > z \ge z_{\rm vir}\\
      \xi R_{\rm vir}
      & {\rm if\ } z < z_{\rm vir}\\
  \end{array}}\right .
\end{equation}
where $z_{\rm ta} \simeq 1.5(1+z_{\rm vir})-1$ is the turn-around
redshift, $t(z)$ is the time corresponding to redshift $z$, and the DM
density inside the halo before $z_{\rm ta}$ is given by equations
(23)\footnote{In the original reference (T97) this formula is affected
by a typografical error, as the sign inside the exponential is wrong.}
and (24) of T97, which are a good approximation of the top hat results:
\begin{eqnarray}
\rho_{\rm TH}(z) & = & 
\rho_0 \Omega_{\rm DM} (1+z)^3 \left[{1 +
    \left({e^{{1.9A}\over{1-0.75A^2}}-1}\right)}\right],
\label{onezone_density_evolution}\\
A & = &
A(z)\equiv (1+z_{\rm vir})/(1+z).
\end{eqnarray}
Finally, $\xi$ (for which we will consider values in the range
$0.01-0.2$) is the ratio between the final size of the flat density core
and the virial radius, which we define as
\begin{equation}
R_{\rm vir}\equiv {1\over2} R_{\rm tr}(z_{\rm ta}) =
{1\over2}\left({{3\over{4\pi}}
{{M_{\rm DM}}\over{\rho_{\rm TH}(z_{\rm ta})}}}\right)^{1/3}.
\end{equation}
Such a definition is unusual, but the difference with the most
common definitions of the virial radius (\eg Padamanbhan 1993) is less
than 0.4 per cent.

The above equations divide the evolution of the DM profile in three
stages: before the halo turn-around, it is reasonable to assume that the
density profile within the halo is flat (therefore, $R_{\rm core}=R_{\rm
tr}$). After the turn-around the density profile is evolved smoothly,
reaching an equilibrium configuration at the virialization redshift;
after that, the DM profile becomes completely static.


This is a crude model of the dark matter halo, but it is adequate for
our purposes.  We emphasize that the choice of a model with a {\it flat}
core, rather than a cusp as in the NFW profile (Navarro, Frenk \& White
1997), helps to ensure that the behaviour we observe near the centre is
due to the self-gravity and hydrodynamics of the {\it simulated} gas,
rather than to the {\it assumed} DM profile. However, in the following
we will also discuss the effects of a cuspy NFW profile for the static
post-virialization phase.

\begin{table}
\caption{Assumed chemical composition at z=1000}
\label{initial_chemical_abundances}
\begin{tabular}{lcl}
\hline\hline
Species(X) & $n_{\rm X}/n_{\rm H}\,^{a}$ & Comments\\
\hline
H$^0$     & $0.9328$              & $<1$ because H is partly ionized\\
H$^+$     & $0.0672$              & Sasaki \& Takahara 1993$^{b}$\\
H$^-$     & $10^{-19}$            & GP98, fig. 4\\
\HH       & $10^{-13}$            & GP98, fig. 4\\
H$_2^+$   & $10^{-18}$            & GP98, fig. 4\\
He        & $0.0833$              & \\
He$^+$    & $10^{-25}$            & GP98, fig. 4\\
He$^{++}$ & $0$                   & \\
$e^-$     & $0.0672$              & from charge conservation\\
D$^0$     & $2.332\times 10^{-5}$ & (Romano \etal 2003)$^{c}$\\
D$^+$     & $1.68\times 10^{-6}$  & (Romano \etal 2003)$^{c}$\\
HD        & $2.5\times 10^{-18}$  & (Romano \etal 2003)$^{c}$\\
\hline\hline
\end{tabular}
\\
$^{a}$ $n_{\rm H}$ is the {\it total} abundance of H,
including all species, \ie $n_{\rm H} = n_{\rm H^0} + n_{\rm H^+} +
n_{\rm H^-} + 2(n_{\rm H_2} + n_{\rm H_2^+}) + n_{\rm HD}$.\\
$^{b}$ from their model with $\Omega_0=1$, $\Omega_{\rm b}=0.05$,
$h=0.5$ (tab. 1); this value also agrees with the result of the RECFAST code
(Seager, Sasselov \& Scott 1999, 2000) for the flat $\Lambda$CDM
cosmology we are assuming.\\
$^{c}$ Romano \etal (2003) actually give the {\it total} abundance of
D, $n_{\rm D} = 2.5\times10^{-5} n_{\rm H}$; we split this into the
D$^0$, D$^+$ and HD abundances by assuming that $n_{\rm D^+}/n_{\rm
D}=n_{\rm H^+}/n_{\rm H}$, and that $n_{\rm HD}/n_{\rm D}=n_{\rm
H_2}/n_{\rm H}$.
\end{table}

\subsection{The models}


%

\subsubsection{Initial conditions}

We start our computations at a very high redshift ($z=1000$), when both
the dark matter and baryonic overdensities are still very small, and it
is perfectly reasonable to assume an uniform distribution of the gas
(the effects of Compton drag further improve the accuracy of this
assumption - see eq. 4.171 of Padmanabhan 1993). Furthermore, we
simulate a region whose comoving initial radius is 10 times larger than
that of the halo, in order to achieve a better modeling of the
hydrodynamical effects, as this guarantees that the total mass in the
simulated region is well above both the cosmological Jeans mass and the
filtering mass (see \eg\ Peebles 1993, and Gnedin 2000).

We assumed that at $z=1000$ the gas temperature is the same as that of
the CMB ($\simeq 2728\;{\rm K}$), that its density is equal to the
cosmological average for baryons ($\rho_{\rm gas}(z=1000) \simeq
4.22\times 10^{-22}\; {\rm g\,cm^{-2}}$), and that the chemical
abundances are those listed in Table \ref{initial_chemical_abundances}.

\subsubsection{Simulation sets}

We ran several sets of simulations (listed in
Table~\ref{list_simulation_sets}), covering halos with virialization
redshifts in the range $10\leq z_{\rm vir}\leq 100$ and total
(baryonic+dark matter) halo masses in the range $5\times10^3\msun\leq
M_{\rm halo}\leq M_{\rm H}(z_{\rm vir})$. $M_{\rm H}$ is the minimum
mass of the halos in which the cooling from atomic H is effective, that
is, the minimum mass of halos with $T_{\rm vir} \geq 10^4\;{\rm K}$
\begin{equation}
M_{\rm H}(z_{\rm vir}) \simeq 1.05 \times 10^9 \msun
  \left(\frac{\mu}{1.23}\right)^{-3/2} (1+z_{\rm vir})^{-3/2},
\label{mass_atomic_cooling2}
\end{equation}
where $\mu$ is the mean molecular weight of the primordial gas.

Each simulation set includes {\it two} runs for each $(z_{\rm vir},
M_{\rm halo})$ pair, differing only because HD cooling is either
included or omitted. The results of the runs without HD cooling are used
as a control sample. The models are based on the initial conditions
described above, and consist of 150 shells, whose mass increases from
the centre to the outskirts. The spacing was chosen in order to achieve
sufficient mass ``resolution'' at the centre: the mass of the central
shell is always $\sim 0.3\msun$, which is a very small fraction ($\sim
10^{-3}$) of the mass of the fragments seen in the simulations of ABN02,
BCL02, and Y06.

\begin{table*}
\caption{Summary of simulation sets}
\label{list_simulation_sets}
\begin{flushleft}
\begin{tabular}{llll}
\hline\hline
Set & DM profile & Abundances & Remarks\\
\hline
$\xi=0.1$  & Isothermal$^a$ ($\xi=0.1$)  & from
Table~\ref{initial_chemical_abundances} & Fiducial \\
extra-D    & Isothermal$^a$ ($\xi=0.1$)  & High$^c$ D, e$^-$, H$^+$ &
Doubled rate of reaction 15\\
$\xi=0.2$  & Isothermal$^a$ ($\xi=0.2$)  & from
Table~\ref{initial_chemical_abundances} & \\
$\xi=0.05$ & Isothermal$^a$ ($\xi=0.05$) & from
Table~\ref{initial_chemical_abundances} & \\
$\xi=0.01$ & Isothermal$^a$ ($\xi=0.01$) & from
Table~\ref{initial_chemical_abundances} & \\
NFW        & NFW$^b$ ($c=10$)            & from
Table~\ref{initial_chemical_abundances} & \\
\hline\hline
\end{tabular}
\\
$^{a}$ Truncated isothermal, as described in
section~\ref{subsubsection_darkmatter}.\\
$^{b}$ Before virialization, the DM profile is assumed to evolve as in
the fiducial ($\xi=0.1$) set.\\
$^{c}$ For D, e$^-$, and H$^+$, twice the value given in
Table~\ref{initial_chemical_abundances}. The other abundances given
there are the same, except for H$^0$ (which was reduced to 0.8656 in
order to compensate the increase in H$^+$).
\end{flushleft}
\end{table*}

We ran several sets of simulations, in order to check the effects of the
most uncertain parameters describing our initial conditions, such as the
DM density profile:

\begin{enumerate}
\item{We check the influence of the concentration of the previously
  described truncated isothermal DM density profile by running four sets
  ($\xi=0.2$, $\xi=0.1$, $\xi=0.05$, and $\xi=0.01$) of simulations
  differing for the value of the $\xi$ parameter. We will refer to the
  $\xi=0.1$ set as to the ``fiducial'' set.}
\item{We check whether a cuspy density profile can lead to significantly
  different results by running a set of simulations (NFW) in which the
  post-virialization density profile within $R_{\rm vir}$ is described
  by a NFW provile with concentration parameter $c=10$; in these
  models, the evolution of the DM density profile before virialization
  is the same as for the ``fiducial'' model.}
\item{We check whether HD effects depend strongly on the assumed
  abundances (at $z=1000$) and on the uncertain rate of reaction 15 by
  running a set of simulations (extra-D) in which we double the initial
  values of both the electron (and H$^+$) fraction given in Table
  \ref{initial_chemical_abundances} (to 0.1344), and the deuterium total
  abundance (to $n_{\rm D}/n_{\rm H} = 5\times10^{-5}$); we also assume
  that the rate of reaction 15 is twice as high as its standard
  value. All these changes favour HD formation, and this simulation set
  is mostly useful as an upper limit on the effects of HD (in the case
  that $\xi=0.1$).}
\end{enumerate}

All the runs were stopped when a density $\rho_{\rm coll} = 1.67\times
10^{-19}\;{\rm g\,cm^{-3}}=10^5 m_{\rm H}\,{\rm cm^{-3}}$ is reached at
some shell (generally at the centre), or when the simulation reaches
$z_{\rm stop}=5$, whichever comes first. Our choice of $\rho_{\rm coll}$
ensures that the gas reaches densities where the first run-away collapse
is taking place; furthermore, $\rho_{\rm coll}\gg\rho_{\rm vir}$, which
ensures that the collapse is highly significant, and also that our
results are quite independent of the exact value of $\rho_{\rm coll}$.
Instead, we chose $z_{\rm stop}=5$ because below that redshift the halos
which are cooled by molecules are both cosmologically unimportant, and
extremely unlikely to survive in the unperturbed (and neutral)
conditions we are assuming.

\bfig
\epsfig{figure=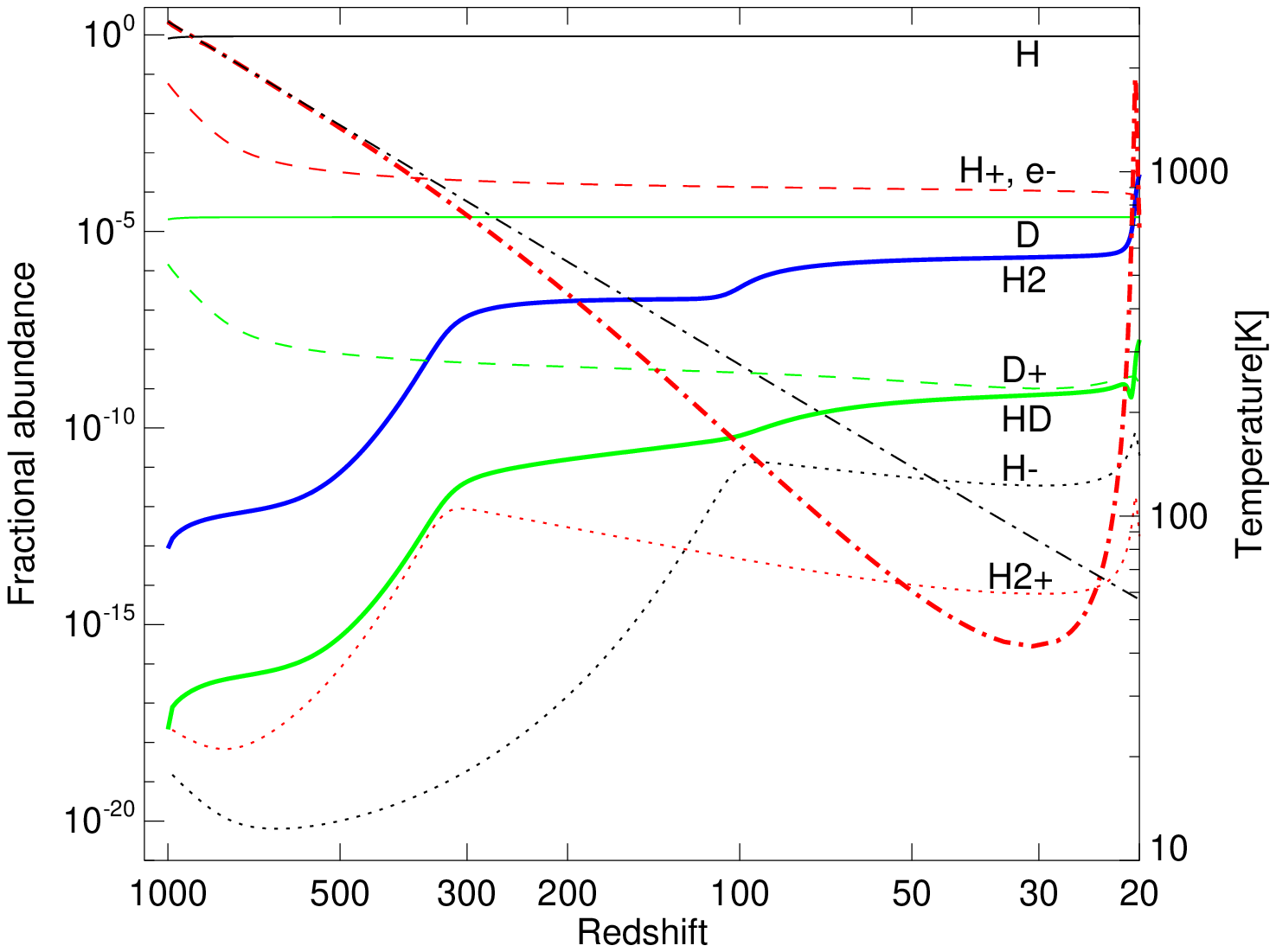,width=8.5truecm}
\caption{Chemical and thermal evolution between recombination and
virialization of a $10^6\msun$ halo with $z_{\rm vir}=20$, taken from
the ``fiducial'' set of simulations. The left axis refers to the labelled
lines, showing the abundance evolution of the various species: the
abundances of H$^+$ and $e^-$ are always extremely close, and are
represented by a single line; neutral and ionized He were omitted. The
right axis refers to the two dot-dashed lines without labels, which show
the evolution of the CMB and of the gas temperature (thin line and thick
line, respectively). The results shown in this plot are almost
independent of $M_{\rm halo}$ and of the inclusion of HD cooling, up to
at least the turn around redshift $z_{\rm ta}=30.5$.}
\label{high_z_chemical_evolution}
\efig

\section{Results}

\bfig
\epsfig{figure=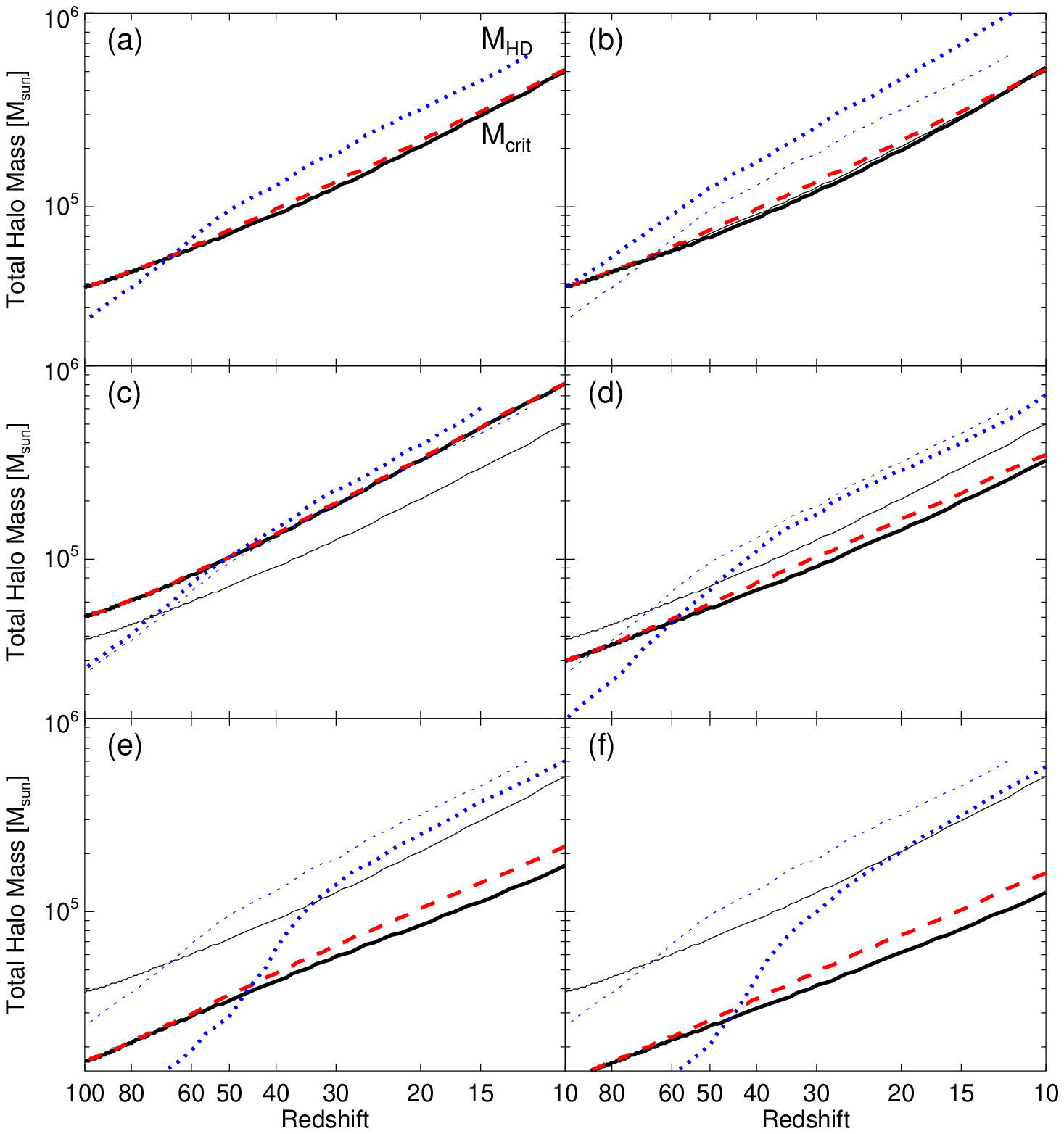,width=8.5truecm}
\caption{Evolution of the critical mass ($M_{\rm crit}$), and of the
  mass below which HD becomes important ($M_{\rm HD}$) as a function of
  the virialization redshift $z_{\rm vir}$, for our six sets of
  simulations. Solid lines show the evolution of $M_{\rm crit}$ when HD
  cooling is included in the simulations, while dashed lines show the
  case where HD cooling is neglected; dotted lines show $M_{\rm
  HD}$. Panels refer to the (fiducial) $\xi=0.1$ set (a), to the extra-D
  set (b), to the $\xi=0.2$ set (c), to the $\xi=0.05$ set (d), to the
  $\xi=0.01$ set (e), and to the NFW set (f). The results of the
  fiducial case are repeated as thin lines in all the panels, in order
  to facilitate the comparison between the various sets.  $M_{\rm crit}$
  is almost unaffected by the inclusion of HD cooling; however, the mass
  range between $M_{\rm HD}$ and $M_{\rm crit}$ becomes significant in
  the extra-D set, or if the DM halos are quite concentrated.}
\label{critical_masses}
\efig

\subsection{Critical masses}

Since the Compton heating due to residual free electrons largely
dominates the thermal behaviour of the ``average'' IGM, the inclusion of
HD cooling scarcely affects the chemical and thermal evolution of the
IGM, even inside regions which will later virialize into halos, such as
those shown in Fig.~\ref{high_z_chemical_evolution}\footnote{Hirata \&
Padmanabhan 2006 have shown that the amount of \HH which formed in the
IGM through the H$_2^+$ and the H$^-$ channels (reactions 3, 5, 9, and
10 of Table~\ref{reactions_list}) at $z\gsim70$ might be substantially
less than what is shown in Fig.~\ref{high_z_chemical_evolution}. In the
rest of this paper we neglect this uncertainty because in our
simulations the bulk of \HH forms inside the halos after they virialize,
and the ``background'' abundance has very little effects on our
results.}.

Instead, HD cooling might affect the thermal evolution of the gas in a
halo before and immediately after virialization, changing the {\it
critical mass} which separates efficiently and inefficiently cooling
halos (see \eg T97).

Each of our simulated halos was classified as {\it collapsing} (or
equivalently, efficiently cooling) or {\it non-collapsing}
(inefficiently cooling), depending on whether it reaches a maximum
density larger than $\rho_{\rm coll}$ in a less than an Hubble time, \ie
at a redshift $z_{\rm coll}\gsim [0.63(1+z_{\rm vir})]-1$.


Efficiently cooling halos have a high probability of forming luminous
objects; this probability is much lower (but sometimes not completely
negligible, as we show in the next section) for inefficiently cooling
halos.

Our definition of collapsing halos is comparable to the ``collapse
criterion'' of T97 (see their \S 5.1), so we use a similar notation and
define $M_{\rm crit}(z_{\rm vir})$ as the minimum mass of an efficiently
cooling halo virializing at $z_{\rm vir}$.

\bfig
\epsfig{figure=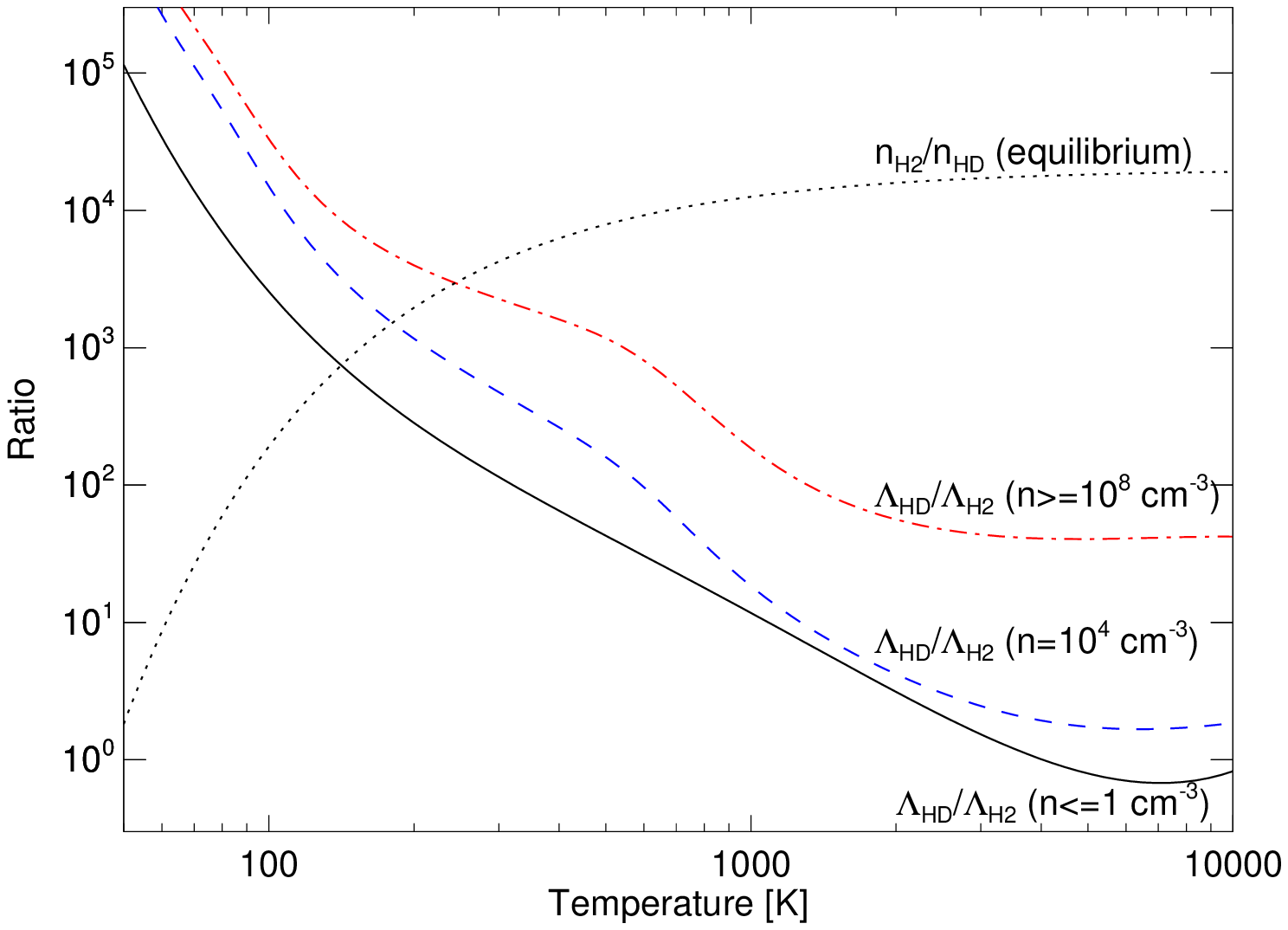,width=8.5truecm}
\caption{Ratio of the HD cooling per molecule to the \HH cooling per
  molecule as a function of temperature at three different densities:
  $1\;{\rm cm^{-3}}$ ({\it i.e.}\ low density limit; solid line),
  $10^4\;{\rm cm^{-3}}$ (dashed line), and $10^8\;{\rm cm^{-3}}$ ({\it
  i.e.}\ high density limit; dot-dashed line). They are compared to the
  $n_{\rm H_2}/n_{\rm HD}$ ratio (dotted line) which can be expected if
  reactions 15 and 16 of Table \ref{reactions_list} are at equilibrium
  (see {\it e.g.}\ of Shchekinov \& Vasiliev 2006): if this assumption
  is correct, the intersections between this line and the other three
  lines mark the conditions where \HH and HD are equally important
  coolants.}
\label{ratio_lambda}
\efig

In figure \ref{critical_masses} we show the behaviour of $M_{\rm
crit}(z_{\rm vir})$,
comparing the results of the runs including HD cooling with those of the
``control'' runs. All the six sets of simulations are shown. In all
cases HD has very little effect upon $M_{\rm crit}$, as the solid and
dashed curves of each simulation set are always close: models where the
DM is quite concentrated exhibit slighly larger differences; but they
never amount to more than $20-30\%$, even in the case of a NFW profile.


The reason of this result is that HD cooling can overcome \HH cooling
only when the temperature is quite low ($\lsim200\;{\rm K}$, see
fig. \ref{ratio_lambda}). But the gas temperature at virialization is
$T\simeq T_{\rm vir}\sim 1000\;{\rm K}$, and HD can become important
only {\it after} \HH has lowered the gas temperature by a substantial
factor, \ie only if \HH cooling is efficient in the first place.  Such
an explanation is confirmed by Figures \ref{halo_evolution_z20_1e5} and
\ref{halo_evolution_z20_1e6}: before virialization HD cooling might even
exceed \HH cooling, but remains negligible when compared to the
effects of Compton scattering on free electrons; at virialization \HH
overcomes both the Compton (because the increase in density favours \HH
formation, reducing the number of free electrons at the same time) and
the HD contribution (because the increase in temperature reduces the
ratio $\Lambda_{\rm HD}/\Lambda_{\rm H_2}$, and also the ratio $n_{\rm
HD}/n_{\rm H_2}$). So, HD cooling may become dominant only well after
virialization, when \HH has already reduced the temperature by a factor
$\gsim 2$.

\subsection{Fragmentation}

The simulations clearly show that even if HD has very little effect on
{\it which} halos cool, collapse, and form stars, it can affect {\it
how} this happens. Actually, there are objects where HD cooling has no
effect, and objects where the thermal evolution of the gas is deeply
affected by HD.

In Figures \ref{halo_evolution_z20_1e5} and \ref{halo_evolution_z20_1e6}
we show the evolution of the central properties in two halos chosen from
the fiducial set. These halos have the same $z_{\rm vir}$ (20) but
different mass ($M_{\rm halo} = 2\times10^5,\, 10^6\,\msun$). In each
figure the results of runs with and without HD cooling are compared. In
particular, we are interested in the evolution of the Jeans mass, which
is taken to be (cfr. Peebles 1993)
\begin{eqnarray}
M_{\rm J}(T,\rho,\mu) & = & {\pi\over6}
\left({{\pi k_{\rm B} T}\over{G \mu m_{\rm p}}}\right)^{3/2}
\rho^{-1/2}\nonumber\\
& \simeq & 50 \msun T^{3/2} \mu^{-3/2} n^{-1/2}.
\end{eqnarray}

In the case of the smaller halo (fig. \ref{halo_evolution_z20_1e5}) the
final stages of evolution differ, because HD cooling becomes dominant,
anticipating the collapse (in this case $\rho_{\rm coll}$ is reached at
$z\simeq 11.5$ rather than $z\simeq10.3$), and, most importantly,
lowering the gas temperature. In fact, the final temperature is
$\sim 70\;{\rm K}$, \ie 3-4 times lower than in the case where HD was
neglected; as the Jeans mass is reduced by a factor $\sim 10$, this
behaviour strongly hints towards a change in the fragmentation
properties of the gas inside an halo of this kind. On the other hand, in
the case of the larger halo (fig. \ref{halo_evolution_z20_1e6}), the
presence of HD makes no difference.

\bfig
\epsfig{figure=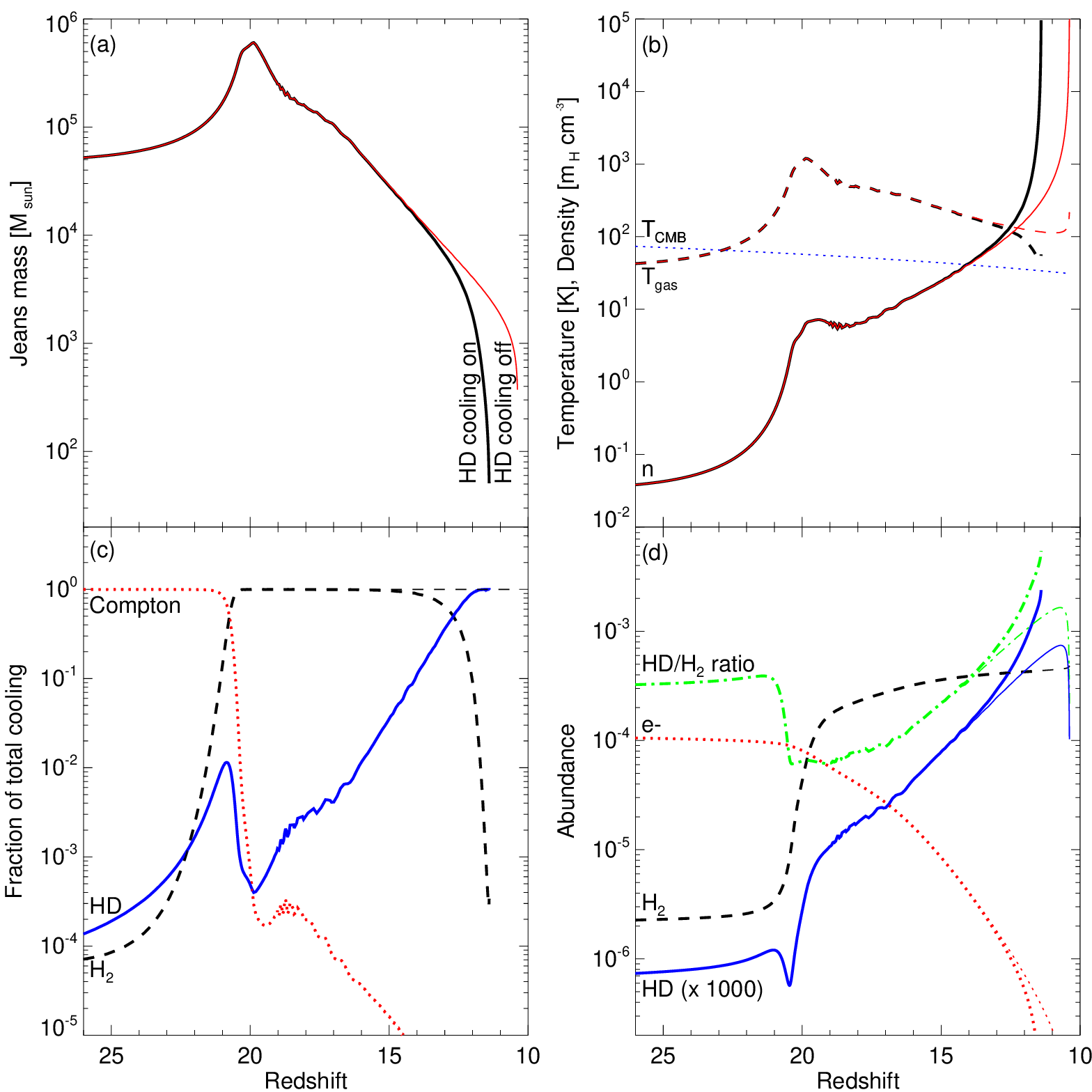,width=8.5truecm,clip=}
\caption{Evolution of the properties of the gas at the centre of an halo
  with $M_{\rm halo}= 2\times10^5\msun$ virializing at $z_{\rm vir}=20$
  (taken from the fiducial set), as a function of redshift. In all
  panels, thick lines refer to models where HD cooling was included,
  whereas thin lines refer to models neglecting HD cooling. The
  properties shown are: Jeans mass (panel a); gas density and
  temperature (panel b; solid and dashed lines, respectively); fraction
  of the total cooling/heating due to HD, \HH and Compton scattering
  (panel c; solid, dashed and dotted lines, respectively); fractional
  abundance of HD, \HH and electrons (panel d; solid, dashed and dotted
  lines, respectively; the HD abundance shown here is $10^3$ times
  larger the real value). In panel (b) the value of the CMB temperature
  is also shown (dotted line), and in panel (d) we additionally show the
  evolution of the $n_{\rm HD}/n_{\rm H_2}$ ratio (dot-dashed line).}
\label{halo_evolution_z20_1e5}
\efig

Figures \ref{halo_evolution_z20_1e5} and \ref{halo_evolution_z20_1e6}
are illustrative of a general trend: HD is unimportant in the largest
halos, but it becomes relevant at lower masses.
In fact, when the largest halos are considered, the \HH abundance just
after virialization is above the threshold ($\sim 5\times 10^{-4}$)
identified by previous studies (\eg T97) for \HH cooling to be
efficient. As \HH sheds away most of the compressional heating, the
collapse proceeds unimpeded; the gas remains at temperatures $T$ above
the $200\;{\rm K}$ ``threshold'' below which HD cooling becomes
important ($T$ usually reaches a minimum between $200$ and $300\;{\rm
K}$, then slowly increases as the collapse proceeds). Instead, in
smaller halos the \HH abundance is below the threshold for efficiently
dispersing the compressional heating, and the gas undergoes a stage of
slow contraction and cooling; for example, in
fig.~\ref{halo_evolution_z20_1e5} this phase lasts from $z\sim20$ to
$z\sim13$.  The ``extra'' time spent in this phase allows a build up of
HD (see especially panel (d) of fig.~\ref{halo_evolution_z20_1e5}),
until HD completely dominates the cooling and is finally able to
``restart'' the collapse.

In figure \ref{critical_masses} we compare the critical mass $M_{\rm
crit}(z)$ to the mass $M_{\rm HD}(z)$ below which HD cooling becomes
important ($M_{\rm HD}(z)$ was taken as the maximum mass where the
inclusion of HD cooling takes the final gas temperature below $200\;{\rm
K}$, with a reduction of at least 25 per cent when compared to the value
obtained when HD is neglected), for all the simulation sets we
considered. The range between $M_{\rm HD}$ and $M_{\rm crit}$ (\ie the
collapsing halos where HD is likely to be important) is negligibly
narrow for $\xi=0.2$, but it gets wider when an higher D abundance is
considered, or when the DM halos are assumed to be more concentrated
than in the fiducial case (see also Section \ref{discussion_section}).

It is important to remark that our results are not in contradiction with
the detailed simulations of BCL99, BCL02, and Y06, because all of them
considered halos with masses in the range $0.5-1\times10^6\msun$ at
redshift $\sim 20$, \ie above $M_{\rm HD}$. We must also remark that our
results depend on the assumption that halos remain unperturbed, because
mergers lead to dynamical heating, which could prevent the collapse also
in halos with mass larger than $M_{\rm crit}$ (see Yoshida \etal 2003;
Reed \etal 2005). Since the probability of mergers is particularly high
in halos where HD is important (because of the long ``build-up'' phase),
this effect is likely to reduce the fraction of halos where HD cooling
is important (see also Section \ref{mass_fraction_section}).

\bfig
\epsfig{figure=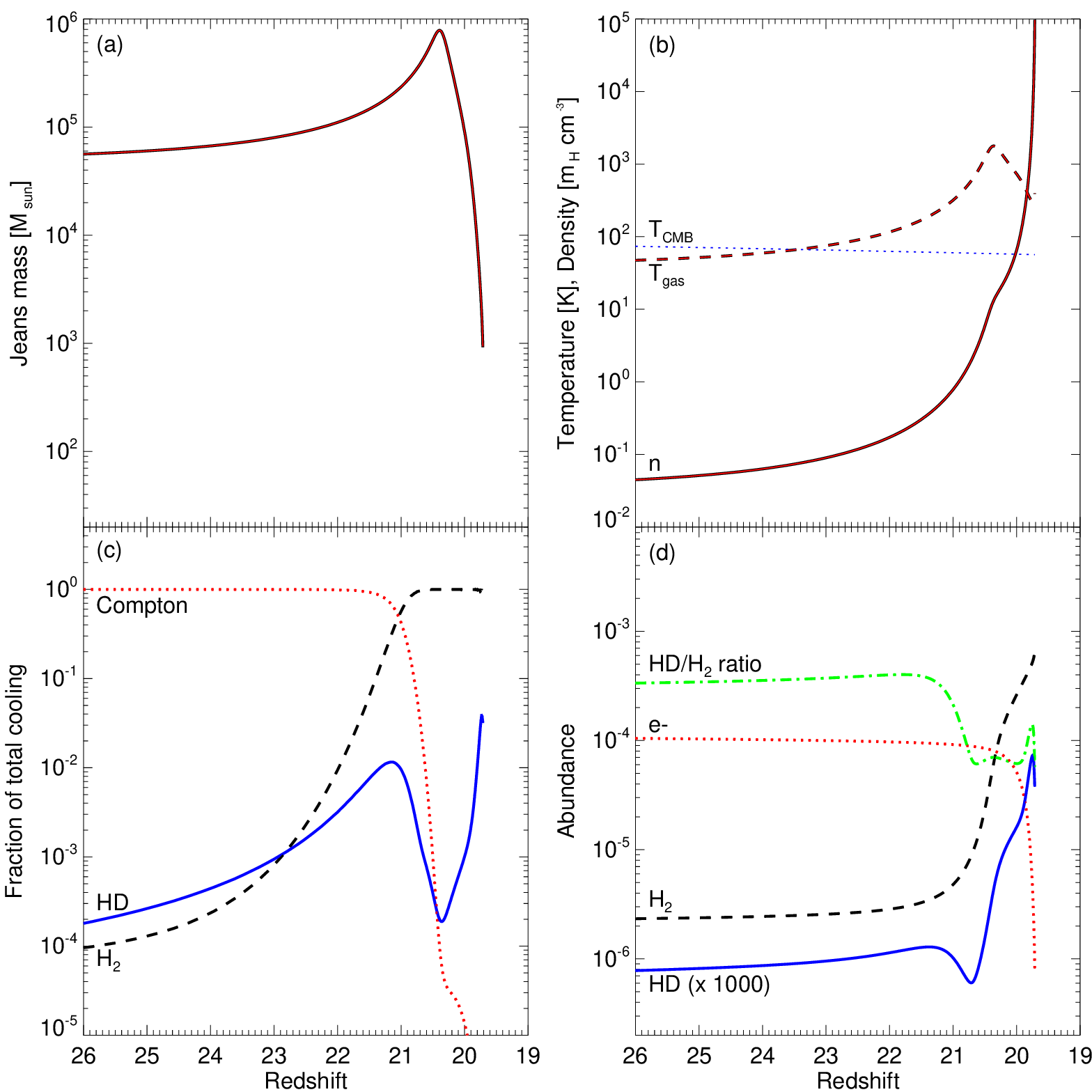,width=8.5truecm,clip=}
\caption{Evolution of the properties of the gas at the centre of an halo
  with $M_{\rm halo}\simeq 10^6\msun$ virializing at $z_{\rm vir}=20$,
  (taken from the fiducial set) as a function of redshift. See the
  caption of fig. \ref{halo_evolution_z20_1e5} for a description of the
  panels and an explanation of the symbols. The absence of the thin
  lines associated with the case where HD cooling was not included is
  only apparent: in halos as massive as the one shown here, the gas
  evolution is not influenced by HD cooling, and the thin and thick
  tracks are perfectly superimposed.}
\label{halo_evolution_z20_1e6}
\efig

\subsection{Proto-stellar collapse}

Lipovka \etal (2005) suggested that HD cooling might be relevant also at
relatively high densities and temperatures ($T\lsim 3000\;{\rm K}$,
$n\gsim10^6$), a regime which is associated with proto-stellar collapse.
Another reason to check this regime is that HD cooling might be in the
optically thin regime even when the \HH cooling is substantially reduced
by radiative transfer effects.

In order to evaluate these effects, we re-ran some of our models with a
better resolution (200 shells, mass of the central shell $\sim
10^{-3}\,\msun$) stopping them only after a density $n=10^{13}\;{\rm
cm^{-3}}$ was reached (after that, the fast increase of CIE cooling will
overwhelm both \HH and HD line cooling, see \eg Ripamonti \& Abel
2004). We explored 6 virialization redshifts ($z_{\rm vir}=10,\; 15,\;
20,\; 30,\; 50,\; 100$), always choosing an halo mass of $10^6\msun$,
above both $M_{\rm crit}(z)$ and $M_{\rm HD}(z)$): here we ignore halos
where HD can affect fragmentation, because otherwise we cannot have a
meaningful comparison with a model where HD cooling was neglected.

Again, by comparing simulations with and without HD cooling, (see \eg
fig. \ref{proto_evolution_z30_1e6}) we find that HD causes no difference
also in this phase of the formation of a primordial star. The main
reason is that the ratio $n_{\rm HD}/n_{\rm H_2}$ decreases with
temperature (at equilibrium, $n_{\rm HD}/n_{\rm H_2} \propto e^{-{465
{\rm K}}\over T}$, see Shchekinov \& Vasiliev 2006), and so does the
ratio of HD cooling per molecule to \HH cooling per
molecule. Furthermore, radiative transfer effects become important only
when the \HH fraction is already of the order of 1, implying $n_{\rm
HD}/n_{\rm H_2}\lsim 2n_{\rm D}/n_{\rm H}\simeq 5\times10^{-5}$
($10^{-4}$ if the abundances of the ``extra-D'' set are considered):
this difference is simply too large to be compensated by the radiative
transfer effects on \HH cooling efficiency, especially at the relatively
high temperatures ($\sim 1000\;{\rm K}$) typically associated with the
high-density phases of the collapse.

We emphasize that the above calculations always assumed the optically
thin limit for HD cooling. We have not checked whether this is correct,
but our results will not change even if it is not: in fact, the
treatment of radiative transfer would then yield an effective HD cooling rate
which is {\it lower} than what we assumed.

\bfig
\epsfig{figure=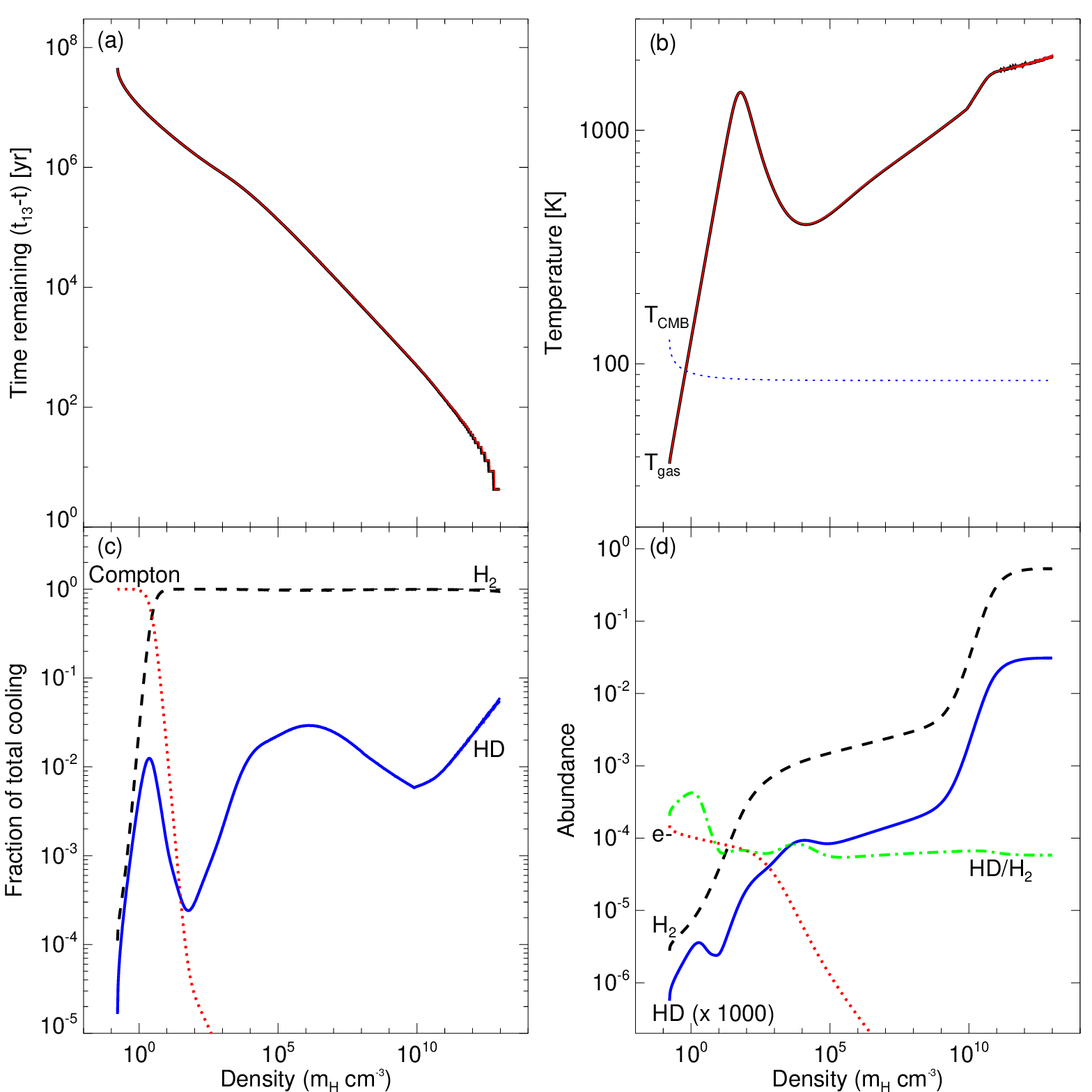,width=8.5truecm,clip=}
\caption{Evolution of the properties of the gas at the centre of an halo
  with $M_{\rm halo}\simeq 10^6\msun$ virializing at $z_{\rm vir}=30$
  (taken from the fiducial set), as a function of central
  density. Panels (c) and (d) are the same as in
  figs. \ref{halo_evolution_z20_1e5} and
  \ref{halo_evolution_z20_1e6}, except that the X-axis shows the gas central
  density, rather than the redshift. Panel (a) shows the time at which each
  density is reached, expressed as the time ($t_{\rm 13}-t$) remaining
  before reaching a density $10^{13}\;m_{\rm H}\,{\rm cm^{-3}}$ at time
  $t_{\rm 13}$. Panel (b) shows the temperature evolution of the gas
  (solid lines) and of the CMB (dotted line).  As in
  fig. \ref{halo_evolution_z20_1e6}, the absence of the thin lines
  (corresponding to the case where HD cooling was not considered) is
  only apparent: the results with and without HD are almost
  identical. Finally, the increase in the fraction of total cooling due
  to HD at high densities (panel c) is due to radiative transfer effects
  reducing the \HH cooling efficiency; such effects are not considered
  for HD, so the high density part of the solid curve in panel (c)
  should be considered an upper limit.}
\label{proto_evolution_z30_1e6}
\efig

\section{Discussion}\label{discussion_section}

In the previous section we showed that in the standard $\Lambda$CDM
model the main effect that HD might exert upon the evolution of
unperturbed primordial halos is to reduce the typical stellar
mass in objects below a certain mass $M_{\rm HD}(z_{\rm vir})$. Here, we
investigate whether this ``HD-dominated'' regime of star formation is
cosmologically relevant or not. This is done by comparing the mass of
the stars which formed in halos affected by HD cooling, and in halos
where only \HH (or H) cooling was important.

\subsection{Rates of formation and survival of objects}

We base our estimates on a variation of the Press-Schechter formalism
(Press, Schechter 1974), namely the extended Press-Schechter (EPS)
formulae of Sheth \& Tormen (1999) [ST99], which was found to be in
reasonable agreement with the results of numerical simulations of the
formation of very early objects by Gao \etal (2005; see also Jenkins
\etal 2001, Reed \etal 2003, Gao \etal 2004, Springel \etal 2005).

Such formulae can be treated in the same way as Sasaki (1994) did for
the original Press-Schechter, leading to similar results: the comoving
number density of virialized halos in the mass range between $m$ and
$m+dm$ at redshift $z$ is
\begin{eqnarray}
N_{\rm ST}(m,z) & = & \left[{ {{\rho_0}\over m}
\left({{2a}\over\pi}\right)^{1/2}
\left({-{1\over{\sigma(m)^2}}{{d\sigma}\over{dm}}}\right)
A\delta_{\rm c}}\right]\nonumber\\
& & \times\, {{1+(a\nu)^{-p}}\over{D(z)}} e^{-{{a\nu}\over2}}
\label{n_eps}
\end{eqnarray}
where $a=0.707$, $A=0.322$ and $p=0.3$ are the parameters given by ST99
(instead, $a=1$, $A=0.5$, $p=0$ correspond to a ``standard'' Press
Schechter function), $\sigma(m)$ is the value of the root mean squared
fluctuation in spheres that on average contain a mass $m$, $\delta_{\rm
c}\simeq 1.686$ is the overdensity threshold for the collapse, $D(z)$ is
the growth function of perturbations (see Peebles 1993), and
\begin{equation}
\nu(m,z)\equiv {{\delta_{\rm c}^2}\over{D(z)^2 \sigma(m)^2}}.
\label{nu_definition}
\end{equation}

The time derivative of eq. (\ref{n_eps}) is
\begin{eqnarray}
\dot{N}_{\rm ST}(m,z) & = & - {{dD}\over{dz}} {{dz}\over{dt}} {1\over D}
\left({1-a\nu-2p{{(a\nu)^{-p}}\over{1+(a\nu)^{-p}}}}\right)\nonumber\\
& & \times\, N_{\rm ST}(m,z)
\label{ndot_eps}
\end{eqnarray}
and by following Sasaki 1994 (in particular in the assumption that the
destruction rate of halos has no characteristic mass scale) we get that
the formation rate is given by
\begin{equation}
\dot{N}_{\rm form}(m,z) = {{dD}\over{dz}} {{dz}\over{dt}} {1\over D}
{{a\nu+(a\nu)^{1-p}-2p}\over{1+(a\nu)^{-p}}} N_{\rm ST}(m,z)
\label{halo_formation_rate}
\end{equation}
and also that the probability that an object which exists at $z'$
survives until redshift $z$ ($z<z'$) without merging is
\begin{equation}
p(z',z) = \left[{{D(z')}\over{D(z)}}\right]^{1-2p} \simeq
\left[{{1+z}\over{1+z'}}\right]^{1-2p}
\label{survival_probability}
\end{equation}
where the last equality is strictly valid only in an Einstein-de Sitter
universe with $\Omega_{\rm m}=1$.

\subsection{Mass fractions}
\label{mass_fraction_section}

The results of our simulations can be interpolated in order to get an
estimate of the function $z_{\rm coll}(m,z_{\rm vir})$, \ie of the
redshift at which the collapse of the gas inside an halo of mass $m$
that virialized at $z_{\rm vir}$ has proceeded far enough for star
formation to occur; from the same data, we can also estimate a
correlated quantity, \ie the minimum mass $M_{\rm min}(z_{\rm vir},z)$
that an unperturbed halo which virialized at $z_{\rm vir}$ must have in
order to have formed stars before redshift $z$.  These functions can be
combined with the above equations (\ref{halo_formation_rate}) and
(\ref{survival_probability}) in order to find the mass fractions in
object-forming halos of various kinds (HD-cooled, H$_2$-cooled,
H-cooled) as a function of redshift. For example, the fraction of mass
that at redshift z is in halos which have produced (or are producing)
stars through HD-dominated cooling is
\begin{equation}
f_{\rm HD}(z) =
\int_{z_{\rm 1}}^z dz_{\rm vir} {{dt}\over{dz_{\rm vir}}}
\int_{M_{\rm HD,1}}^{M_{\rm HD,2}} dm
{{m}\over{\rho_0}} \dot{N}_{\rm form}(m,z_{\rm vir}) p(z_{\rm vir},z),
\label{HD_halo_mass_fraction}
\end{equation}
where $z_{\rm 1}=100$ is the maximum considered redshift (we assume
that the contribution to the star-forming density of halos from redshift
higher than $z_{\rm 1}$ is negligible), and
\begin{eqnarray}
M_{\rm HD,1} = &
\min (M_{\rm HD}(z_{\rm vir}),M_{\rm min}(z_{\rm vir},z)),\\
M_{\rm HD,2} = &
M_{\rm HD}(z_{\rm vir})
\end{eqnarray}

We note that the $p(z_{\rm vir},z)$ factor in
eq. (\ref{HD_halo_mass_fraction}) excludes all the halos which
experienced a major merger between $z_{\rm vir}$ and $z$; this is a very
approximate way of accounting for the effects of dynamical heating
(Yoshida \etal 2003; Reed \etal 2005). Some
of these excluded halos (namely, the ones undergoing a merger at a
redshift between their $z_{\rm coll}$ and $z$) actually formed stars in
the HD regime. This effect could be included in the calculation by
changing the probability factor into $p(z_{\rm vir},\max(z,z_{\rm
coll}(m,z_{\rm vir})))$. However, the difference between the two
probabilities is relatively small: in the most extreme case (\ie $z_{\rm
vir}=z_{\rm coll} = 100$; $z=10$) $p(z_{vir},z_{\rm
coll})/p(z_{vir},z)\simeq2.41$, whereas for much more typical values of
$z_{\rm vir},\ z_{\rm coll}$, and $z$ the correction factor is $\lsim
1.5$, which is too small to alter the qualitative conclusions that can
be reached through eq.~(\ref{HD_halo_mass_fraction}).

The mass fractions $f_{\rm H_2}(z)$ and $f_{\rm H}(z)$ (due respectively
to objects where \HH cooling and atomic H cooling is dominant) can be
found by simply changing the limits of the mass integration inside
eq. (\ref{HD_halo_mass_fraction}) to
\begin{eqnarray}
M_{\rm H_2,1} = &
\min(M_{\rm H}(z_{\rm vir}),
\max(M_{\rm HD}(z_{\rm vir}),M_{\rm min}(z_{\rm vir},z))),\\
M_{\rm H_2,2} = &
M_{\rm H}(z_{\rm vir})
\end{eqnarray}
and to
\begin{eqnarray}
M_{\rm H,1} = &
\max(M_{\rm H}(z_{\rm vir}), M_{\rm min}(z_{\rm vir},z)),\\
M_{\rm H,2} = &
\infty.
\end{eqnarray}

\bfig
\epsfig{figure=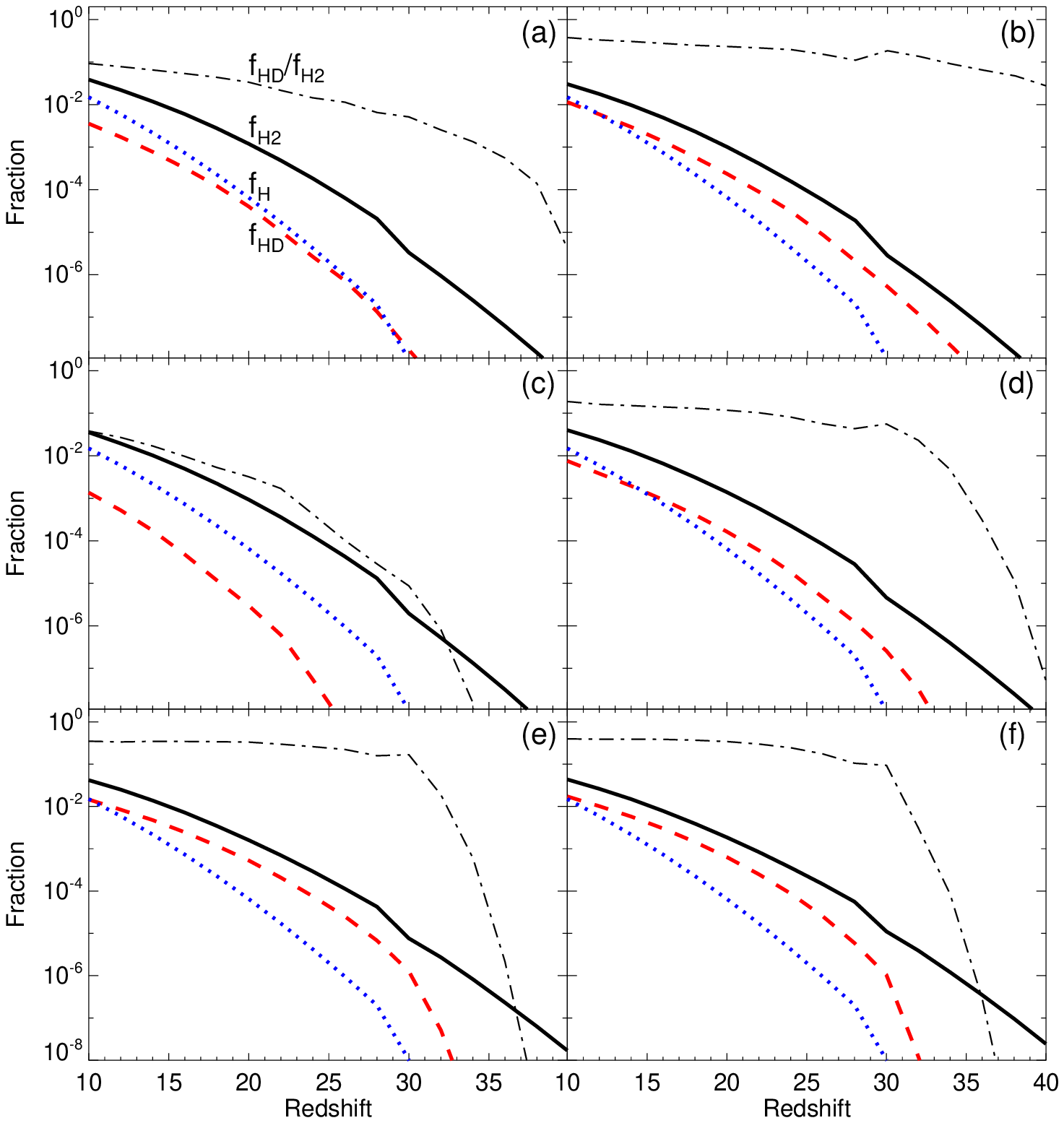,width=8.5truecm,clip=}
\caption{Evolution of the fraction of mass inside halos which were able
  to form stars within a given redshift. Solid lines: mass inside
  \HH-cooled halos; dashed lines: mass inside HD-cooled halos; dotted
  lines: mass inside H-cooled halos; dot-dashed lines: ratio between the
  masses inside HD-cooled halos and \HH-cooled halos. Panels refer to
  different sets of simulations, and are in the same order as in
  Fig.~\ref{critical_masses}.}
\label{sf_fraction}
\efig

We calculated $f_{\rm HD}$, $f_{\rm H_2}$ and $f_{\rm H}$ by taking the
$\sigma(m)$ and $D(t)$ functions given in Eisenstein \& Hu 1998. In
fig.~\ref{sf_fraction} we compare their redshift evolution. It is
apparent that $f_{\rm HD}<f_{\rm H_2}$ at all redshifts and in all the
cases we are considering. However, there is a substantial difference
between the various simulation sets. In the least concentrated case
($\xi=0.2$) the amount of star formation through the HD cooling regime
is likely negligible, never exceeding 4 per cent of the total and rapidly
decreasing when going to high redshifts. Instead, when the DM profiles
are assumed to be moderately or highly concentrated ($\xi\lsim0.05$, or
NFW profile), the difference between the HD and the \HH channels for
star formation reduces to a factor $\sim 3$; it is also essentially
constant between $z=10$ and $z=30$, even if the efficiency of the HD
channel rapidly drops at $z\gsim 30$ (a behaviour which could be
expected from fig.~\ref{critical_masses}); the fiducial set is somewhat
in the middle, while the results from the extra-D set resemble the ones we
obtained for moderately concentrated DM profiles, except that the drop
in $f_{\rm HD}/f_{\rm H_2}$ is much slower.

\section{Summary and conclusions}
In this paper we examined the possible effects of HD
cooling upon the most popular scenarios for the formation of first
objects.

We found that HD cooling has very little effect upon the evolution of a
halo before and during virialization, so that the critical mass $M_{\rm
crit}$ separating star-forming and non-star-forming halos is scarcely
affected.
Similarly, we found that HD is even less important when the first phases
of protostellar collapses are considered.
Both these conclusions are quite independent of the shape of DM
profiles, and of the exact abundance of Deuterium.

The most interesting of our results is that HD can influence the
fragmentation process of primordial gas clouds into stars. In fact, our
simulations show that in halos with masses just above $M_{\rm crit}$, HD
cooling dominates the phases of gas collapse when fragmentation is
likely to take place, significantly reducing the gas temperature, the
Jeans mass, and probably also the typical mass of fragments.  If so, HD
cooling could open a channel for the formation of relatively low mass
stars even in metal-free gas.

We then employed a simple analytical model, based on the extended
Press-Schechter formalism, in order to estimate the importance of this
``HD mode'' of star formation. A comparison with the common
``H$_2$-only'' mode of primordial star formation shows that the HD mode
is always sub-dominant. However, its importance depends on the detailed
DM density profile, on D abundance, and on the rate of HD formation. If
DM profiles are assumed to be relatively loose, and if we take the
``standard'' values for D abundance and HD formation rate (as in our
simulation sets $\xi=0.2$ and $\xi=0.1$), the range of halo masses where
HD is important is quite narrow, and the number of stars whose formation
is triggered by HD cooling is probably negligible.  Instead, in the case
of quite concentrated profiles, and/or of higher D abundances and HD
formation rate (such as in our $\xi=0.05$, $\xi=0.01$, NFW, and extra-D
simulations sets), HD dominated halos should account for about one
quarter of all primordial star formation, at least if we are not
underestimating too much the effects of dynamical heating.

We remark that there exist at least two mechanisms which could boost the
importance of HD for primordial star formation. First of all, it is not
unreasonable to expect that the star formation efficiency is higher for
the halos which form stars through HD than for those where HD is
unimportant. This is because the lower typical stellar mass should imply a
weaker feedback.

A second possibility comes from considering that the ratio of HD to \HH
(and with it the mass range where HD cooling is important) is quite
enhanced when the abundance of free electrons is higher than in the
standard case (see \eg Nagakura \& Omukai 2005, O'Shea \etal 2005,
Johnson \& Bromm 2006, Yoshida 2006, even if all of them consider a
different scenario). This might happen because of DM annihilations or
decays (Ripamonti, Mapelli \& Ferrara 2006), or, less exotically,
because of the influence of nearby ionizing sources (\eg accreting black
holes, as in Zaroubi \& Silk 2005).

\section{Aknowledgements}

The author thanks M. Mapelli, T. Abel, S. Zaroubi, E. Vasiliev, and
M. Colpi for useful discussions in the preparation of this paper, and
aknowledges support from the Netherlands Organization for Scientific
Research (NWO) under project number 436016.

\label{lastpage}
\end{document}